\documentclass[%
aip,
 amsmath,amssymb,
 reprint,%
]{revtex4-1}

\usepackage{graphicx}
\usepackage{dcolumn}
\usepackage{bm}
\usepackage[mathlines]{lineno}
\usepackage[utf8]{inputenc}
\usepackage[T1]{fontenc}
\usepackage{mathptmx}
\usepackage{etoolbox}
\usepackage[usenames,dvipsnames,svgnames,table]{xcolor}

\makeatletter
\def\@email#1#2{%
 \endgroup
 \patchcmd{\titleblock@produce}
  {\frontmatter@RRAPformat}
  {\frontmatter@RRAPformat{\produce@RRAP{*#1\href{mailto:#2}{#2}}}\frontmatter@RRAPformat}
  {}{}
}%
\makeatother

\begin{document}

\title{Particle dispersion and clustering in surface ocean turbulence with ageostrophic dynamics}

\author{Michael Maalouly}
\email[]{michael.maalouly@univ-lille.fr}
\affiliation{Univ. Lille, ULR 7512, Unité de Mécanique de Lille Joseph Boussinesq (UML), F-59000 Lille, France}

\author{Guillaume Lapeyre}
\affiliation{LMD/IPSL, CNRS, Ecole Normale Supérieure, PSL Research University, 75005 Paris, France}

\author{Bastien Cozian}
\affiliation{LMD/IPSL, CNRS, Ecole Normale Supérieure, PSL Research University, 75005 Paris, France}
\affiliation{Université Claude Bernard Lyon 1, 69100 Villeurbanne, France}

\author{Gilmar Mompean}
\affiliation{Univ. Lille, ULR 7512, Unité de Mécanique de Lille Joseph Boussinesq (UML), F-59000 Lille, France}

\author{Stefano Berti}
\affiliation{Univ. Lille, ULR 7512, Unité de Mécanique de Lille Joseph Boussinesq (UML), F-59000 Lille, France}

\date{\today}

\begin{abstract}
Upper-ocean turbulent flows at horizontal length scales smaller than the deformation radius 
depart from geostrophic equilibrium and develop important vertical velocities, which are key to marine ecology and climatic processes.
Due to their small size and fast temporal evolution, these fine scales are difficult to measure during oceanographic campaigns.
Instruments such as Lagrangian drifters
have provided another way to characterize these scales through the analysis of pair-dispersion 
evolution, and have pointed out 
striking particle convergence events.  
By means of numerical simulations, we investigate such processes in a model of surface-ocean turbulence
that includes ageostrophic motions.  
This model originates from a Rossby-number expansion of the primitive equations and reduces to the surface quasi-geostrophic model, a paradigm of submesoscale dynamics, in the limit of vanishing Rossby number.
We focus on the effect of the ageostrophic dynamics on the pair-dispersion and clustering properties of Lagrangian tracer particles at the ocean surface.
Our results indicate that while over long times the pair separation process is barely affected by 
the ageostrophic component of the velocity field, the latter is responsible for the formation of temporary particle aggregates, and the intensity of this phenomenon increases with the Rossby number. 
We further show that Lagrangian tracers preferentially accumulate in cyclonic frontal regions, 
which is in agreement with observations and other more realistic modeling studies.
These findings appear interesting to improve the understanding of the turbulent transport by ocean fine scales, 
and in light of upcoming, new high-resolution satellite data of surface velocity fields. 
\end{abstract}



\maketitle


\section{Introduction}
\label{sec:intro}

Ocean flows at scales comparable and smaller than the deformation radius, i.e. in the meso and submesoscale ranges, 
are characterized by quasi two-dimensional (2D) turbulent dynamics. 
In spite of this important common feature, remarkable differences distinguish submesoscales from mesoscales. 
Flow structures in the mesoscale range have horizontal sizes of several tens to few hundreds of kilometers
and they extend over depths of $O(1000)$~m. Such eddies contain most of the kinetic energy in the ocean.
Their vertical velocities, however, are quite small, namely of $O(1-10)$~m~day$^{-1}$.
On the other hand, submesoscales correspond to eddies and, importantly, filaments with smaller 
horizontal scales of $O(1-10)$~km. 
These structures reach depths of only $O(100)$~m, and evolve on faster timescales of $O(1)$~day. 
Theoretical arguments and high-resolution numerical simulations indicate that their vertical velocities 
can be up to an order of magnitude larger than the mesoscale ones~\cite{McWilliams2016,ZQKT2019}. 
They are then expected to provide a relevant contribution to vertical transport, and thus to play a key role for 
both marine ecology and the coupling between the ocean and the atmosphere \cite{Kleinetal2019}.

In recent years, many evidences about submesoscales have emerged from Lagrangian drifter data. Based on the possibility to relate particle pair-dispersion statistics to the properties of the underlying turbulent flow (see, e.g., Ref.~\onlinecite{LaCasce2008}), several authors focused on the determination of the laws controlling the spreading process of drifters deployed at the surface of the ocean. 
By taking this approach, and computing the scale-by-scale pair separation rate, regimes of enhanced relative dispersion at fine scales were detected in different regions, pointing to energetic submesoscales (see, e.g., Refs.~\onlinecite{LE2010,BDLV2011,Poje_etal_2014,CLPSZ2017}).

Another striking feature that was recently observed, first in the Gulf of Mexico~\cite{Dasaro_etal_2018} and later in other regions, is the occurrence of temporary drifter clustering.
This means that while globally Lagrangian particles still spread in time, every now and then many of them are brought together in
regions of very limited size. 
Such convergence events are associated with large vorticity (and divergence) values highlighting the departure from geostrophic balance - meaning that the Rossby number, roughly estimated by $Ro=\zeta/f$ (with $\zeta$ relative vorticity and 
$f$ Coriolis frequency), is not negligibly small - and with the onset of important vertical velocities.

Explaining this phenomenon is currently an open point, and requires going beyond the quasi-geostrophic (QG) approximation, obtained from a development of the basic equations of motion (primitive equations) at the lowest order in $Ro$, in which the flow is strictly horizontal and non-divergent.
To include the physics of both clustering and dispersion, a natural possibility is to improve the dynamics of an idealized QG model by adding higher-order corrections when developing (in $Ro$) the primitive equations. 
While, by construction, the resulting model does not include important sources of ageostrophy, such as high-frequency motions (internal gravity waves and tides), which are further off from geostrophic equilibrium, it properly accounts for ageostrophic motions associated with frontogenesis. 
Moreover, it allows separating the geostrophic and ageostrophic flow components in a straightforward manner.

Understanding the role of ageostrophic turbulent dynamics on Lagrangian transport is relevant in view of  future satellite measurements, such as those from the Surface Water and Ocean Topography (SWOT) mission. 
This satellite, launched at the end of 2022, has started measuring sea surface height (SSH) at a spatial resolution of $\approx 15$~km, 
which represents an order of magnitude of improvement with respect to presently available data~\cite{Morrow_etal_2019}. 
As a result, it should provide access to the fine mesoscale and submesoscale ranges at global scale. 
Determining to what extent small-scale processes, associated with non-negligible Rossby numbers, hinder the possibility to retrieve surface currents from SSH through geostrophic balance represents an important challenge for the exploitation and the theoretical interpretation of these new data.
For this purpose, Lagrangian statistics based on drifter datasets appear promising; different from Eulerian ones, they reflect the temporal evolution of fluid parcels, and may thus enable a clear separation between fast (ageostrophic) processes, that could contaminate the satellite-derived velocity, and slower (geostrophic) ones.

In this study, by means of numerical simulations, we investigate the spreading of Lagrangian tracer particles at the ocean surface in a model of upper-ocean turbulence derived as an extension of the QG approximation, and including ageostrophic effects. 
We particularly focus on the reproduction of Lagrangian convergence events, and on the quantification of the importance of the latter with increasing Rossby number.
Furthermore, by comparing pair-dispersion statistics for particles advected by flows at different values of $Ro$, we aim at assessing the relevance of ageostrophic motions on the relative dispersion process.

This article is organized as follows. In Sec.~\ref{sec:model} we introduce the flow model; the main 
features of its turbulent dynamics are discussed in Sec.~\ref{sec:turb}. The results of the analysis of Lagrangian 
particle statistics are reported in Sec.~\ref{sec:lagr}. There, we separately characterize the role of ageostrophic motions on relative dispersion (Sec.~\ref{sec:reldisp}), and the clustering properties, as well as their relation with the flow structure (Sec.~\ref{sec:cluster}). Finally, discussions and conclusions are presented in Sec.~\ref{sec:concl}.

\section{Model}
\label{sec:model}

A convenient theoretical framework to address the dynamics of the upper ocean in the fine-scale range
(scales comparable and, to some extent, smaller than the deformation radius) is offered by QG models. 
Indeed, these models allowed a relatively good understanding of the larger 
mesoscale [$O(100)$~km] regime~\cite{McWilliams2016}, and can be taken as the basis for model improvement when approaching the lower end [$<O(10)$~km] of the fine-scale range.
They are obtained from an expansion at lowest order in $Ro$ of the momentum and buoyancy evolution equations, 
within the Boussinesq and hydrostatic approximations (see, e.g., Ref.~\onlinecite{Vallis2017}). 
The main dynamical equation, resulting from this approach, assumes constant stratification and states that in the interior of the considered fluid layer 
potential vorticity (PV) is conserved along the geostrophic flow.
 
Surface quasi-geostrophy (SQG)~\cite{HPGS1995,Lapeyre2017} is a special case of QG dynamics. Within this model
the interior PV is assumed to be exactly equal to zero. The associated flow is then entirely driven by
the evolution of surface buoyancy (or, equivalently, temperature). Previous studies highlighted the interest of this model for ocean submesoscale turbulence (see Ref.~\onlinecite{Lapeyre2017} for a review), for phytoplankton diversity~\cite{Perruche_etal11} as well as Lagrangian dispersion \cite{FBPL2017,valade_2023}. 
Indeed, SQG dynamics give rise to energetic small-scale flows,  
and are considered as one of the possible mechanisms of submesoscale generation via mesoscale straining processes. While other mechanisms can also be invoked, such as mixed-layer instabilities, which energize submesoscales also at depth and can be related to the seasonal cycle~\cite{CFFF2016,BL2021}, the SQG model presents the advantage of a simpler mathematical formulation.

Observations, as well as realistic or primitive-equation-based simulations, however, 
revealed some important features, such as the asymmetry of vorticity statistics, with cyclones prevailing over anticyclones~\cite{Rudnick2001,RK2010,Shcherbina_etal_2013}, and the occurrence of Lagrangian convergence events~\cite{HLJK2015,Jacobs_etal_2016,Dasaro_etal_2018,Berta_etal_2020}, which cannot be explained by QG theory. 
In order to overcome the limitations of the QG framework, an interesting possibility is to extend 
it by including ageostrophic motions through the development of primitive equations to next order in $Ro$. By doing so, one obtains the QG$^{+1}$ system, which 
encompasses ageostrophic corrections~\cite{MSR1999,Weiss2022}, potentially responsible of those phenomena. 
In the case of surface-driven dynamics, this approach leads to the so-called SQG$^{+1}$ model. The latter was first introduced in an atmospheric context in Ref.~\onlinecite{HSM2002}, where it was shown through simulations of freely decaying turbulence that it gives rise to the expected cyclone-anticyclone asymmetry.

Here we consider the SQG$^{+1}$ system to investigate surface-ocean turbulence in the fine-scale range, 
a question that to our knowledge has not been addressed before. Our main aim is to provide a minimal model, 
based on the fundamental dynamical equations, accounting for the above mentioned submesoscale features, 
and to use it to investigate the effect of the ageostrophic flow on the spatial distribution of tracer particles.
Other models based on a Rossby-number development of primitive equations exist, such as the surface semi-geostrophic one \cite{ragone_badin_2016}, which reproduces both cyclone-anticyclone asymmetries and strong vertical velocities at fronts. Here we chose the SQG$^{+1}$ model as several of its properties have been well documented.

In the following we shortly introduce the mathematical formulation of the model, adapting the original derivation (see Ref.~\onlinecite{HSM2002} for more details) to the present oceanic conditions. 
We assume that the vertical coordinate is $-\infty < z \leq 0$, and that the dynamics are controlled by the lateral advection of temperature (buoyancy) at the surface ($z=0$). The main governing equation retains the same form as in the SQG system (corresponding to $Ro=0$), and it expresses the conservation of surface temperature along the surface flow. This reads: 
\begin{equation}
\partial_t \theta^{\scriptsize (s)} + \bm{u}^{\scriptsize (s)} \cdot \bm{\nabla} \theta^{\scriptsize (s)} = 0,
\label{eq:sqgp1_temp}
\end{equation}
where $\theta(\bm{x},t)$ is the temperature fluctuation field, the superscript $(s)$ indicates quantities evaluated at $z=0$, and the total velocity field is given by the sum of the geostrophic component $\bm{u}_g$ (computed at the lowest order in $Ro$) and two (next order in $Ro$) ageostrophic terms $\bm{u}_\varphi$ and $\bm{u}_a$,
\begin{equation}
\bm{u}=\bm{u}_g + Ro \left( \bm{u}_\varphi + \bm{u}_a \right).
\label{eq:sqgp1_veltot}
\end{equation}
The geostrophic velocity can be expressed in terms of the streamfunction $\phi$: 
\begin{equation}
\bm{u}_g=\left( -\partial_y \phi, \partial_x \phi \right),
\label{eq:sqqp1_u0}
\end{equation}
where $x$ and $y$ denote the horizontal coordinates.  Note that here and in what follows we use nondimensional units. 
As in SQG, the streamfunction is related to surface temperature through 
\begin{equation}
\phi = \mathcal{F}^{-1}\left[ \frac{\mathcal{F}(\theta^{\scriptsize (s)})}{k}e^{kz} \right],
\label{eq:sqgp1_phi_cap}
\end{equation}
where $\theta$ is here taken at lowest order, $\mathcal{F}$ stands for the horizontal Fourier transform and $k$ for the horizontal wavenumber modulus.  
The above relation is a direct consequence of the assumption of zero interior PV, 
$\bm{\nabla}^2_H \phi +\partial_z^2 \phi=0$ (with $\bm{\nabla}^2$ the Laplacian operator and the subscript $H$ indicating that only horizontal coordinates are considered), with the boundary conditions $\theta^{\scriptsize (s)}=\partial_z \phi|_{z=0}$  and $\partial_z \phi \rightarrow 0$ for $z \rightarrow -\infty$. 
The ageostrophic velocity components, absent in SQG, can be expressed as 
\begin{equation}
\bm{u}_\varphi=\left( -\partial_y \varphi, \partial_x \varphi \right),
\label{eq:sqqp1_uphi}
\end{equation}
\begin{equation}
\bm{u}_a=-\partial_z \bm{A},
\label{eq:sqgp1_ua}
\end{equation}
where the functions $\varphi$ and $\bm{A}$
are related to surface and lower-order quantities by:
\begin{equation}
\varphi = \frac{\theta^2}{2} - \mathcal{F}^{-1}\left\{ \frac{\mathcal{F}\left[\theta^{\scriptsize (s)} 
(\partial_z \theta)^{\scriptsize (s)} \right]}{k}e^{kz} \right\},
\label{eq:sqgp1_phi}
\end{equation}
\begin{equation}
\bm{A} = -\theta \bm{u}_g + \mathcal{F}^{-1}\left[ \mathcal{F}(\theta^{\scriptsize (s)} \bm{u}_g^{\scriptsize (s)}) e^{kz} \right],
\label{eq:sqgp1_A}
\end{equation}
again with $\theta$ taken at lowest order. Equation~(\ref{eq:sqgp1_phi}) follows from the requirement of having zero interior PV at all orders in $Ro$, while Eq.~(\ref{eq:sqgp1_A}) is a form of the omega equation obeyed by vertical velocities (see also Refs.~\onlinecite{MSR1999,Lapeyre2017,HSM2002}). 
The functions $\varphi$ and $\bm{A}$ are such that $\partial_z\varphi=0$ and $\bm{A}=\bm{0}$ at $z=0$.
Note that $\bm{u}_a$ has both a rotational and a divergent component from (\ref{eq:sqgp1_A}) while $\bm{u}_{\varphi}$ is nondivergent. 

Remark that the model specified by Eqs.~(\ref{eq:sqgp1_temp})-(\ref{eq:sqgp1_A}), by construction,  
accounts for ageostrophic motions related to fronts, meaning those associated with next-order corrections to the balanced (i.e. geostrophic) flow. Other sources of ageostrophy are instead excluded. 
In particular, this applies to higher-frequency motions, such as internal gravity waves and tides, which are not close to geostrophic equilibrium.

\section{Turbulent flow properties}
\label{sec:turb}

The model evolution equations (Sec.~\ref{sec:model}) are numerically integrated by means of a pseudospectral method 
on a doubly periodic square domain of side $L_0=2\pi$ at resolution $N^2=1024^2$, starting from an initial condition corresponding to a streamfunction whose Fourier modes have random phases and small amplitudes. 
The code was adapted from an original one developed by Ref.~\onlinecite{Smith_etal_2002} and previously used 
in Refs.~\onlinecite{BL2014,FBPL2017,BL2021}. 
We consider the forced and dissipated version of Eq.~(\ref{eq:sqgp1_temp}), which allows reaching a statistically stationary flow state. 
Specifically, we add on the right-hand side of the equation a random ($\delta$-correlated in time) forcing acting 
over a narrow range of wavenumbers $4\leq k_f \leq 6$ (and whose intensity is $F=0.02$), as well as a hypofriction term $-\alpha \bm{\nabla}_H^{-2} \theta$ to remove energy from the largest scales, and a hyperdiffusion term $-\nu\bm{\nabla}_H^4\theta$ to assure small-scale dissipation and numerical stability. 
For the dissipative terms we set  $\alpha=0.5$ and we determine $\nu$ according to the condition $k_{max} l_\nu \gtrsim 6$, with $l_\nu$ the dissipative scale (estimated for $Ro=0$). 
These choices correspond to quite large dissipations, and will limit the number of active scales; however, it turned out that they were necessary for controlling  the numerical stability of the code at the largest $Ro$ value explored. Indeed, the integration of the SQG$^{+1}$ system is delicate due to the effective compressibility of the horizontal flow introduced by the ageostrophic corrections, which creates strong gradients that are difficult to resolve. The surface-temperature evolution equation, Eq.~(\ref{eq:sqgp1_temp}) with forcing and dissipation terms, is advanced in time using a third-order Adams-Bashforth scheme. 
We verified that the results are essentially unchanged when using a fourth-order Runge-Kutta algorithm, 
but the latter is computationally less efficient. The time step was set to the quite small value $dt=10^{-4}$, 
which was verified to ensure temporally converged results for different values of the Rossby number. 
The latter being the main control parameter, we performed different simulations by increasing it from $Ro=0$ to $Ro=0.075$, which is the largest value we can safely reach.   

In the following we present the main characteristics of the turbulent flows, for both $Ro=0$ (SQG) and $Ro>0$ (SQG$^{+1}$) that will be of interest for the dynamics of Lagrangian tracer particles. 

\subsection{Kinetic energy spectra}
\label{sec:ke_spectra}
 
When the Rossby number is increased, starting from $Ro=0$, the flow develops stronger and stronger gradients  
and the total kinetic energy grows monotonically with $Ro$ (not shown). 
Its spatial structure is characterized by eddies of different sizes and, especially, by sharp fronts (see also Sec.~\ref{sec:lagr}).  

\begin{figure}
    \includegraphics[scale=0.68]{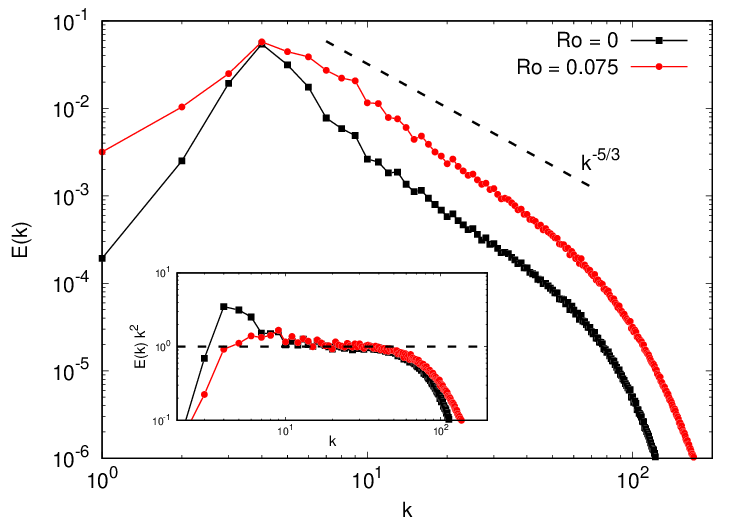}
    \caption{Kinetic energy spectra, temporally averaged over several flow realizations in the statistically steady state for $Ro = 0$ and $Ro=0.075$. The dashed black line in the main panel corresponds to the expectation for SQG dynamics.
    Inset: the same spectra compensated by $k^{-2}$ and rescaled with a coefficient such that,
    in both cases, the scaling range corresponds to the wavenumbers for which $E(k)k^{2} \simeq 1 $.}
    \label{fig:f1}
\end{figure}

Kinetic energy spectra $E(k)$ computed from the total velocity $\bm{u}$, for the smallest ($Ro=0$) and the largest ($Ro=0.075$) Rossby number are shown 
in Fig.~\ref{fig:f1}. They display a scaling close to $k^{-2}$ (see inset of Fig.~{\protect\ref{fig:f1}}) over about a decade. They are flatter than in QG barotropic dynamics, where $E(k) \sim k^{-3}$. However they are slightly steeper than the theoretical prediction $k^{-5/3}$ for the direct cascade of buoyancy variance in the SQG system. 
This steepening effect is essentially independent of $Ro$ and is more important at low wavenumbers, suggesting that its origin likely lies in the presence of large-scale persistent structures of size $\approx 2\pi/k_f$, as also noted in previous studies of SQG and SQG$^{+1}$ turbulence~\cite{CCMV2004,BL2014,Lapeyre2017,HSM2002}.

At high wavenumbers the scaling range is limited by the large values of the dissipation coefficients, which are needed to control the formation of very intense gradients. At low wavenumbers, we do not observe the $k^{-1}$ scaling corresponding to an inverse cascade in SQG, as the forcing acts on large scales and hypofriction is strong enough to damp modes below $k_f$.

\subsection{Vorticity statistics}
\label{sec:vortstat}

As mentioned earlier, an important feature of oceanic (and atmospheric) flows, which is not captured by QG theory, is the asymmetry of vorticity statistics. This was detected in data from both observations~\cite{Rudnick2001,Shcherbina_etal_2013}  and primitive-equation simulations~\cite{LKH2006,RK2010}. The latter numerical works also highlighted the role of surface dynamics on the prevalence of cyclonic over anticyclonic flow regions. 

Different mechanisms can explain this asymmetry. A first one is related to nonlinear Ekman pumping. As the 
stress at the air-sea interface is proportional to the difference of winds and currents, it creates a surface drag causing the decay of ocean anticyclones~\cite{dewar87,zavala_sanson_etal_2023}.
Another mechanism relies on the vortex-stretching term
in the vorticity equation 
$\partial_t\zeta \sim (f+\zeta)\partial_z w + ...$ for finite Rossby numbers. Here $w$ is the vertical velocity,
$f$ the Coriolis frequency and relative vorticity is defined as $\zeta=\partial_x v - \partial_y u$ [where $\bm{u}=(u,v)$ is the horizontal flow].
As discussed in previous works (see, e.g., Refs.~\onlinecite{HSM2002,MT2006,McWilliams2016}), at fronts, through the ageostrophic term $\zeta \partial_z w$, vortex stretching amplifies more cyclonic vorticity (on the heavy side of the front) than anticyclonic vorticity (on the light side of the front).  
Note that within a purely QG framework vortex stretching would instead give a contribution to the vorticity growth rate 
($\partial_t \zeta \sim f \partial_z w$)
that is independent of the sign of $\zeta$. 

\begin{figure}
    \includegraphics[scale=0.68]{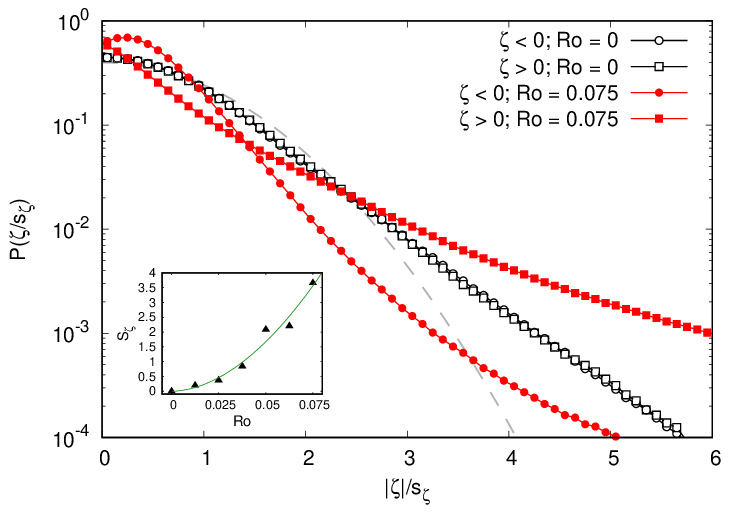}
    \caption{Probability density function of vorticity $\zeta$ (rescaled by its rms value $s_\zeta$), temporally averaged over several flow realizations in the statistically steady state, for $Ro=0$ (empty black points) and $Ro=0.075$ (filled red points), with different  
    point types indicating $\zeta>0$ and $\zeta<0$. 
    For reference, the standard Gaussian distribution is also shown (dashed gray curve).
    Inset: vorticity skewness $S_\zeta$ as a function of the Rossby number; the solid green line corresponds to $S_\zeta \sim Ro^{1.87}$.} 
    \label{fig:f2}
\end{figure}

Clear asymmetry in favor of stronger cyclones is also observed in QG$^{+1}$ and SQG$^{+1}$ models in which next-order corrections in $Ro$ to QG equations are included~\cite{MSR1999,HSM2002}. 
It was argued that the symmetry is broken because the divergence due to ageostrophic frontogenesis at small scales accelerates (slows down) the contraction of dense (light) filaments~\cite{HSM2002,HC2005}, which gives rise to intense and localized cyclones, and weaker more broadly spread anticyclones.
This is the case in our forced simulations of SQG$^{+1}$ turbulence as cyclones prevail over anticyclones whenever $Ro>0$, and vorticity statistics are similar to those in decaying turbulence at fixed Rossby number~\cite{HSM2002}. 
The probability density function (pdf) of $\zeta$, rescaled by its standard deviation $s_\zeta$ and averaged over time, is shown in Fig.~\ref{fig:f2} for $Ro=0$ and $Ro=0.075$. As it can be seen in the figure, the right tail of the pdf ($\zeta>0$) is much higher than the left one ($\zeta<0$) when $Ro=0.075$, while the two tails essentially overlap over a whole range of $|\zeta|$ values for $Ro=0$. The skewness of the vorticity distribution $S_\zeta=\langle \zeta^3 \rangle/\langle \zeta^2 \rangle^{3/2}$ grows, approximately quadratically, with $Ro$ (see inset of Fig.~\ref{fig:f2}), indicating that the magnitude of the asymmetry increases with the intensity of the ageostrophic flow.     

Based on the results in this section, 
the SQG$^{+1}$ simulations considered here appear appealing to explore the transport and dispersion properties of Lagrangian tracers in turbulent flows, relevant for surface-ocean dynamics and possessing (weakly) ageostrophic components.

\section{Lagrangian dynamics}
\label{sec:lagr}

We now consider the dynamics of Lagrangian tracer particles in the turbulent flows produced by the model of Sec.~\ref{sec:model}, both at $Ro=0$ and at $Ro>0$.  
In order to qualitatively compare the main features of our results with those from ocean drifters, we restrict the motion to occur at the surface. Particles then move according to the following equation:
\begin{equation}
    \frac{d\bm{x}_i}{dt}=\bm{u}(\bm{x}_i(t),t),
    \label{eq:motiontracers}
\end{equation}
where $\bm{x}_i=(x_i,y_i)$ is the horizontal position of particle $i$ (with $i=1,...,N_p$) and $\bm{u}(\bm{x}_i,t)$ is the total velocity (i.e. including the ageostrophic component, for $Ro \neq 0$) at its position.

Equation~(\ref{eq:motiontracers}) is numerically integrated using a third-order Adams-Bashforth scheme and bicubic interpolation in space of the velocity field at particle positions~\cite{Hua1994}. Except where explicitly stated, we assume that the particle motion occurs in an infinite domain and use the spatial periodicity of the Eulerian flow to compute the Lagrangian velocities outside the computational box. 
The temporal accuracy of the resulting trajectories was verified by varying the time step, and also according to the Lagrangian acceleration criteria proposed in Ref.~\onlinecite{ZB1990}.
A total of $N_p=49152$ particles are seeded in the turbulent flows once the latter are at a statistically steady state. Their initial positions correspond to a regular arrangement of $M=128 \times 128$ triplets over the entire domain. 
Each triplet forms an isosceles right triangle, with a particle pair along $x$ and one along $y$, 
both of which are characterized by an initial separation $R(0)=\Delta x/2$ (with $\Delta x$ the grid spacing). 
In the following, we introduce the distance between two particles $R(t)=\sqrt{R_x(t)^2+R_y(t)^2}$ [where $R_x(t)$ and $R_y(t)$ are the separations along $x$ and $y$, respectively, at time $t$]. To compute dispersion statistics only original pairs were used, which in our case, amounts to $32768$ pairs. It was verified that 
the pair separation statistics 
do not depend on the initial orientation (along $x$ or $y$ direction) of the pairs. Moreover, provided that enough pairs are chosen, the results are mostly insensitive to their number.

\begin{figure*}
    \includegraphics[scale=0.7]{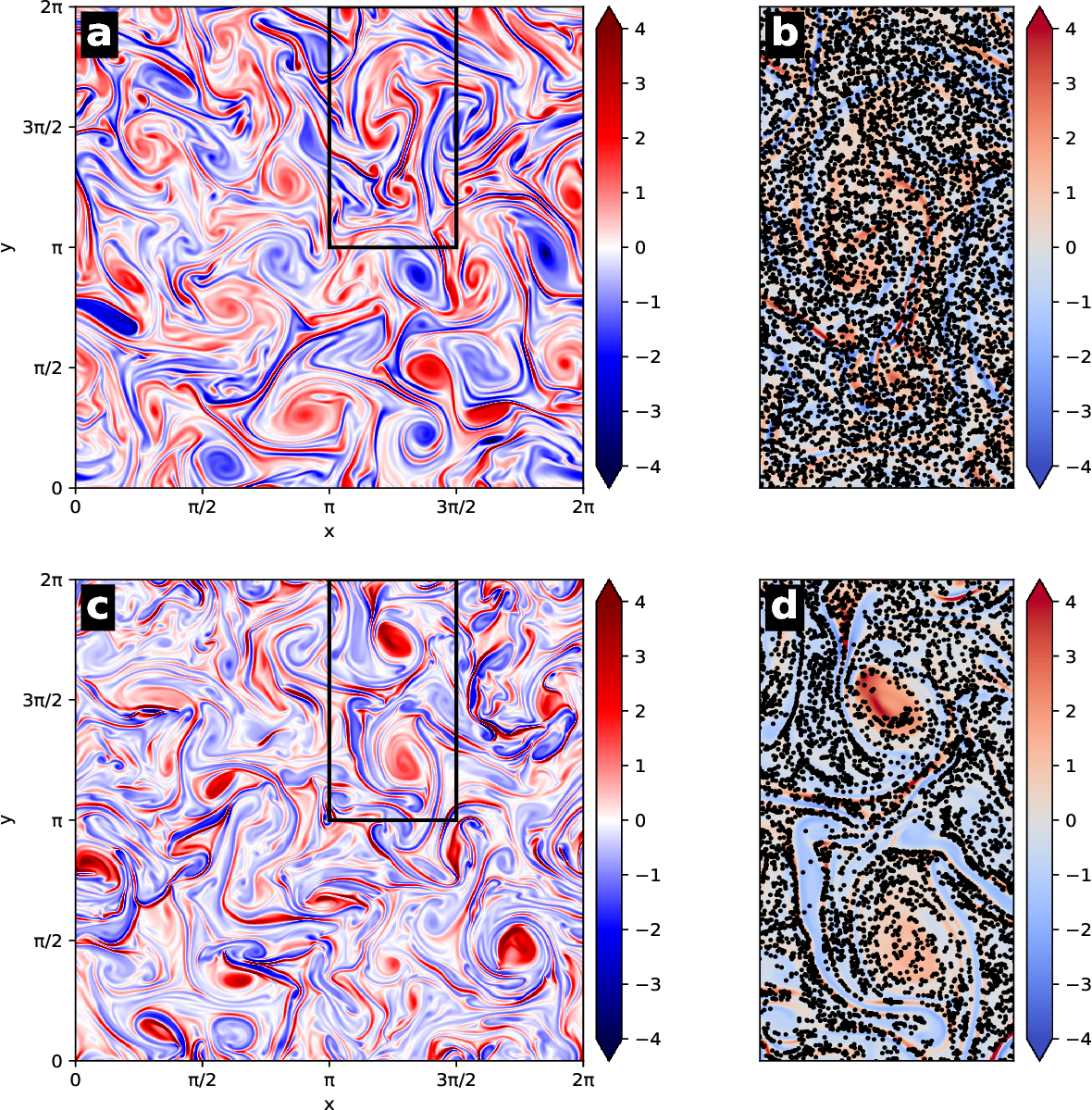}
    \caption{Vorticity normalized by its rms value  
    for $Ro=0$ (a) and $Ro=0.075$ (c) 
    at a fixed instant of time in statistically stationary conditions. Panels (b) and (d) show a
    closeup view of the region in the black rectangle in the main panels (a) and (c), respectively, including the particle distribution at that time. 
    }
    \label{fig:f3}
\end{figure*}

An illustration of typical particle spatial distributions, at a given instant of time in the statistically steady state of the flow, is shown in Fig.~\ref{fig:f3} for both $Ro=0$ and $Ro=0.075$, together with the corresponding vorticity fields. 
Here, particles are placed back in the original doubly periodic domain to see the effect of accumulation in space (while we assume that they leave this domain when computing dispersion statistics).
Independently of the value of $Ro$, vorticity is characterized by quite a filamentary structure in addition to almost elliptical vortices of various sizes. For nonzero $Ro$ cyclonic eddies ($\zeta>0$) are more coherent than anticyclonic ones ($\zeta<0$), and vorticity is globally more intense in root-mean-square (rms) value (not shown).
Concerning particles, it is here apparent that at $Ro=0.075$ they do not uniformly spread 
over the spatial domain (as is the case for $Ro=0$), which highlights the occurrence of clustering. 
In the following, we will separately address the characterization of their relative dispersion process, 
and of their aggregation properties in the flow, for varying Rossby number.

\subsection{Pair-dispersion statistics}
\label{sec:reldisp}

Here, we examine the effect of varying the Rossby number on particle pair dispersion, using both fixed-time 
and fixed-scale indicators. The latter typically better allow to disentangle contributions from different 
flow scales~\cite{ABCCV1997,CV2013,BDLV2011,FBPL2017}. 
We then mainly focus on the scale-by-scale dispersion rate, by computing the finite-size Lyapunov exponent (FSLE)~\cite{ABCCV1997,CV2013}, 
defined as:
\begin{equation}
\lambda(\delta) = \frac{\log{r}}{\langle \tau(\delta) \rangle},
\label{eq:fsle_r}
\end{equation}
where the average is over all pairs 
and $\tau(\delta)$ is the time needed to observe the separation growing from $\delta$ to a scale $r\delta$ (with $r > 1$).

In a nonlocal dispersion regime, for which the separation process 
is controlled by the largest flow features, and normally associated with a steep kinetic energy spectrum of the flow [$E(k) \sim k^{-\beta}$, with $\beta>3$], the FSLE is expected to attain a scale-independent, constant value. 
This reflects in an exponential growth of the mean squared pair separation distance, i.e. relative dispersion: 
\begin{equation}
\langle R^2(t) \rangle = \langle |\bm{x}_i(t) - \bm{x}_j(t)|^2 \rangle.
\label{eq:reldisp}
\end{equation}
Note that relative dispersion is a fixed-time metric, with the average computed at time $t$, over all pairs $(i,j)$ such that at $t=0$ (the release time) $|\bm{x}_i(0) - \bm{x}_j(0)| = R(0)$. 
When the turbulent flow possesses 
energetic small scales [$E(k) \sim k^{-\beta}$, with $\beta<3$], the separation process should be controlled by velocity increments at a length scale comparable to the distance between particles within a pair.
The dispersion regime is therefore referred to as a local one, and both the FSLE and relative dispersion are expected to display power-law behaviors: $\lambda(\delta) \sim \delta^{(\beta-3)/2}$ and $\langle R^2(t) \rangle \sim t^{4/(3-\beta)}$, respectively. At separations larger than the largest flow scales, or at very large times, 
particles in a pair experience essentially uncorrelated velocities and their separation distance   grows diffusively, implying that the FSLE scales as $\lambda(\delta) \sim \delta^{-2}$ and relative dispersion as $\langle R^2(t) \rangle \sim t$. 

Another indicator that may be used to discriminate between different dispersion regimes is the kurtosis of the separation distance:
\begin{equation}
ku(t) = \frac{\langle R^4(t) \rangle}{\langle R^2(t) \rangle^2}.
\label{eq:kurt}
\end{equation}
Under nonlocal dispersion, $ku(t)$ should grow exponentially 
in time, while for local dispersion it should attain a constant value (equal to 5.6 for Richardson dispersion, expected for $\beta=5/3$) at intermediate times~\cite{LaCasce2010,FBPL2017}. 
At very large times, the kurtosis should in any case converge to $ku=2$ corresponding to the diffusive limit of dispersion~\cite{LaCasce2010,FBPL2017}. 

\begin{figure}
    \includegraphics[scale=0.68]{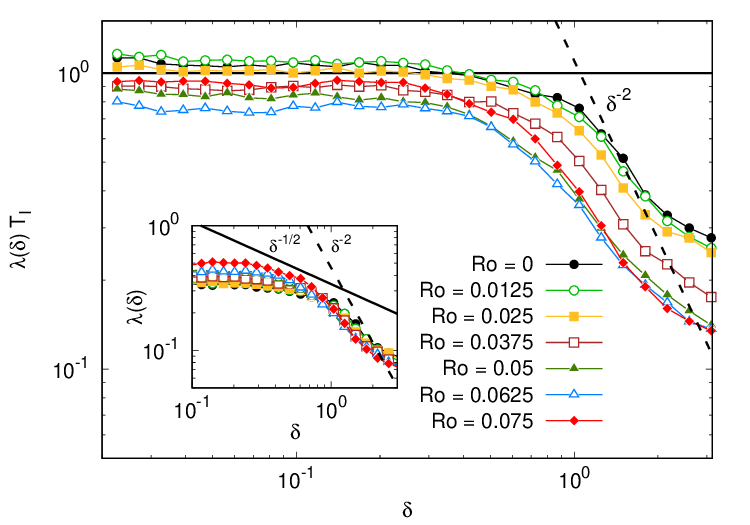}
    \caption{FSLE (rescaled by the flow integral time scale) for different Rossby numbers. Inset: the same without rescaling the FSLE. 
    The $\delta^{-1/2}$ scaling law is the dimensional prediction for a kinetic energy spectrum $E(k) \sim k^{-2}$. 
    The scale amplification factor is $r=1.2$, and it was verified that the results are robust with respect to the choice of this 
    parameter value.}
    \label{fig:f4}
\end{figure}

The FSLE measured in our simulations for different values of the Rossby number is shown in Fig.~\ref{fig:f4}. 
Independently of $Ro$, the curves are remarkably flat at small separations, and approach the diffusive behavior at the largest ones [larger than the flow integral lengthscale $\ell_I = 2 \pi \int_0^\infty k^{-1} E(k) dk/\int_0^\infty E(k) dk$]. 
The slight deviations from the expected $\delta^{-2}$ scaling are here likely due to the limited inertial range  of our turbulent flows. Indeed, previous studies reported similar observations in simulations with reduced inertial ranges, and proposed the use of an alternative, pdf-based indicator~\cite{LM2022} to improve the agreement with the large-scale theoretical prediction.

\begin{figure}
    \includegraphics[scale=0.68]{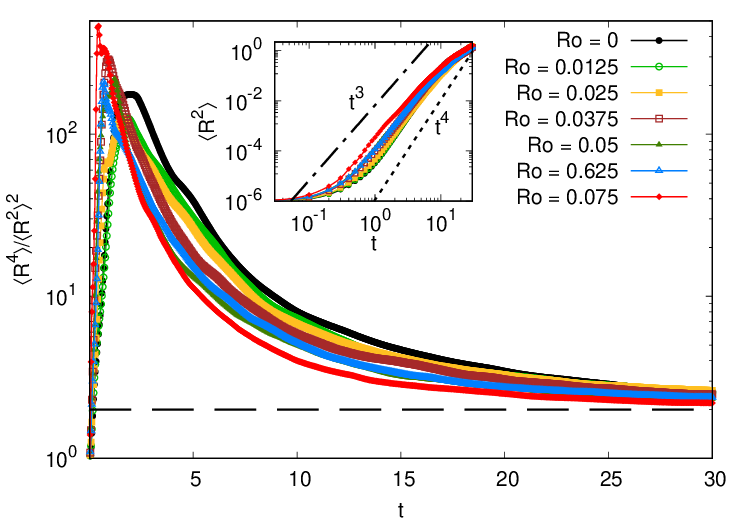}
    \caption{Kurtosis of particle relative displacements (main panel) and relative dispersion (inset) 
    as a function of time for different Rossby numbers. The $t^3$ (Richardson dispersion) and $t^4$ scaling laws in the inset are the expectations for a kinetic energy spectrum $E(k) \sim k^{-5/3}$ and $E(k) \sim k^{-2}$, respectively.}  
    \label{fig:f5}
\end{figure}

No clear evidence of a power-law scaling $\lambda(\delta) \sim \delta^{-1/2}$ [following from a kinetic energy spectrum $E(k) \sim k^{-2}$] is detected, except perhaps on a narrow range of intermediate separations (see inset of Fig.~\ref{fig:f4}). 
This result suggests that the dispersion process is essentially nonlocal. 
This is also confirmed by the temporal evolution of the kurtosis (Fig.~\ref{fig:f5}), which displays a fast growth at short times, and approaches $2$ at large times. 
At intermediate times, 
$ku(t)$ never approaches a constant plateau, which would correspond to a local dispersion regime.
This behavior, pointing to nonlocal dispersion while local dispersion would be expected, may appear quite surprising. 
Interestingly, it bears some resemblance to measurements of drifter separation in the Gulf of Mexico~\cite{Beron2016,sanson2017}, once inertial oscillations are removed. 
One possibility to explain it is related to the presence of large-scale coherent structures in the flow, which can provide a dominant contribution to the dispersion process~\cite{CCMV2004}. 
To test this hypothesis, we rescale the FSLE with the flow integral timescale $T_I=\ell_I/\sqrt{E}$, with $E$ the total kinetic energy. As it can be seen in Fig.~\ref{fig:f4}, 
for all $Ro$, 
the plateau values of the rescaled FSLE range between 1.1 and 0.8, which are close to $1$, supporting this explanation.

The values of FSLE (not rescaled by $T_I$) at small $\delta$ slightly increase with the Rossby number (inset of Fig.~\ref{fig:f4}), consistently with the increase  of velocity gradients with $Ro$.
A similar trend is observed from the short-time behavior of relative dispersion, which grows faster for larger $Ro$ (inset of Fig.~\ref{fig:f5}). 
At later times, $\langle R^2(t) \rangle$ does not present a clear scaling, though on a limited time interval it may not be far from the $t^4$ theoretical expectation. More interestingly, its growth slows down when the Rossby number is increased, which hints to temporary phases during which some particles aggregate and thus the efficiency of the global separation process is reduced.

We conclude that the $Ro$-dependence of the different measures of pair separation is overall weak, indicating that ageostrophic motions do not substantially alter pair-dispersion statistics. This suggests that, in this system, when the Rossby number is increased, large eddies conserve their capacity to drive the dispersion process.

\subsection{Particle clustering and relation with the Eulerian flow structure}
\label{sec:cluster}

While on average, over long times, Lagrangian tracers separate, their spatial distribution 
is not homogeneous and clusters can form in the course of time. To investigate this point, the first quantity 
we consider is the averaged divergence experienced by particles along their trajectories, also known as the dilation rate~\cite{HLJK2015}, a numerically efficient single-particle indicator of tracer accumulation. 
\begin{figure}
    \includegraphics[scale=0.68]{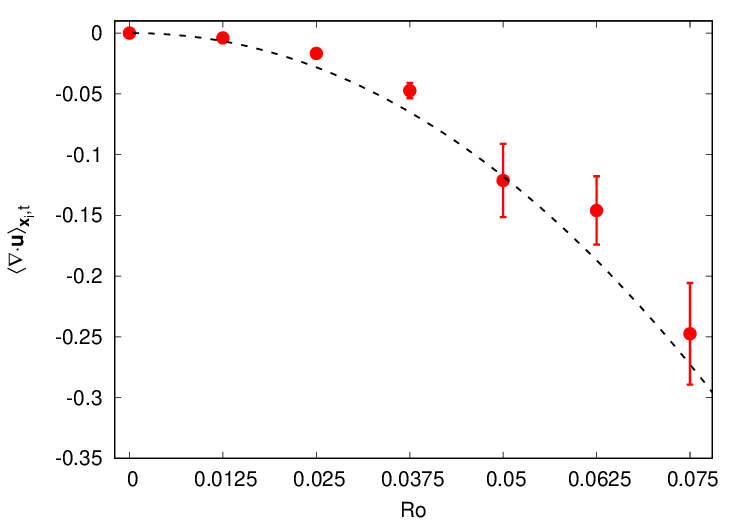}
    \caption{Velocity divergence sampled by particles, averaged over time and over all particles, as a function of the Rossby number. 
    Here the error bars correspond to the standard deviation of the temporal statistics. 
    The black dashed line is proportional to $-Ro^\alpha$, with $\alpha \simeq 2.07$ from a best fit.}
    \label{fig:f6}
\end{figure}

\begin{figure}
    \includegraphics[scale=0.68]{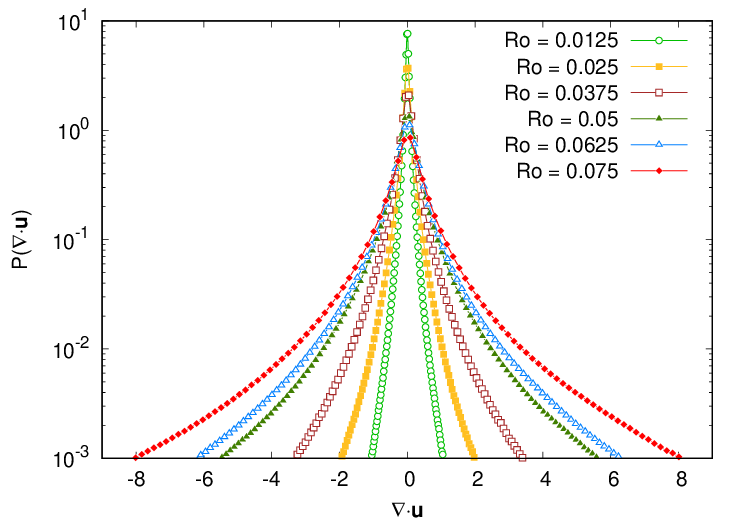}
    \caption{Probability density function of the Eulerian flow divergence $\bm{\nabla} \cdot \bm{u}$, temporally averaged over several flow realizations in the statistically steady state, for different values of $Ro$.}
    \label{fig:f7}
\end{figure}

The divergence of the velocity field 
$\langle \bm{\nabla} \cdot \bm{u} \rangle_{\bm{x}_i,t}$, 
computed at particle positions $\bm{x}_i$ and averaged over time and all particles, 
is shown as a function of $Ro$ in Fig.~\ref{fig:f6}. 
It is negative for nonzero Rossby numbers and grows roughly quadratically in $Ro$ in absolute value, indicating that particles aggregate more when ageostrophic motions are more intense.
Due to the compressibility they experience, particles are attracted to contracting flow regions and hence do not homogeneously sample the phase space. This fact has been shown to give rise to differences between Lagrangian and Eulerian statistics in other situations, such as that of time-correlated compressible flows~\cite{DMV2014,BDES2004}. 
A qualitative understanding on what occurs in our experiments can be obtained by looking at the pdf of the Eulerian divergence, $P(\bm{\nabla} \cdot \bm{u})$ (Fig.~\ref{fig:f7}). When $Ro$ is increased, the tails of this pdf rise, highlighting the more likely occurrence of very intense divergence events. Its shape is remarkably symmetric, though, meaning that positive and negative values of $\bm{\nabla} \cdot \bm{u}$ are equally probable. The negative sign of the averaged Lagrangian divergence $\langle \bm{\nabla} \cdot \bm{u} \rangle_{\bm{x}_i,t}$ then results from particles getting trapped in convergence regions and spending a significant fraction of the time there, a phenomenon which increases in intensity with increasing Rossby number.    

\begin{figure}
     \includegraphics[scale=0.68]{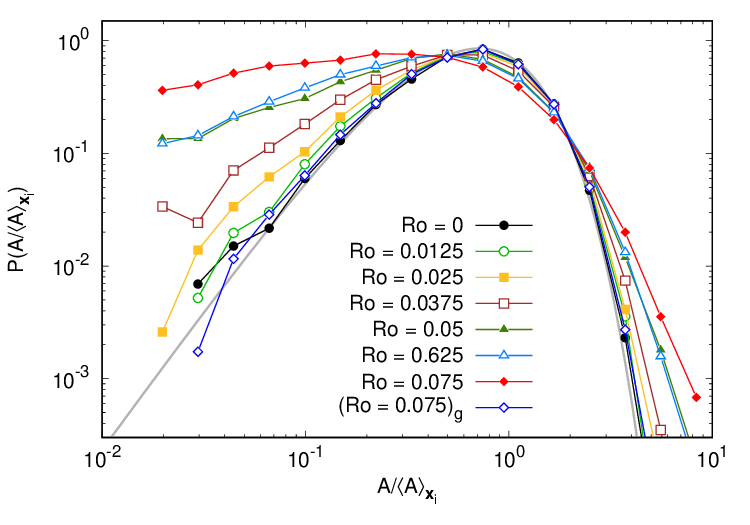}
     \caption{Probability density function of Vorono\"i cell areas, normalized by the averaged cell area, 
     $P(A/\langle A \rangle_{\bm{x}_i})$, at an instant of time in the statistically steady flow state, 
     for different values of the Rossby number. 
     The curve labeled by $(Ro=0.075)_g$ has been obtained from particles advected by the geostrophic flow only. The solid gray line is the theoretical prediction for uniformly distributed particles 
     $f_{2D}\left(A/\langle A \rangle_{\bm{x}_i}\right)$ (see text).}
     \label{fig:f8}
 \end{figure}

The occurrence of clustering in our system is clearly demonstrated by the pdf of Vorono\"i normalized cell areas, a statistical tool that is often used to characterize the aggregation of inertial particles in (incompressible) turbulent flows~\cite{MBC2010,Tagawa_etal_2012}. 
The cells are constructed by partitioning the spatial domain into regions containing one particle and all the points that are closer to that particle than to any other~\cite{Okabe_etal_2000,MBC2010,Tagawa_etal_2012}. 
The nonhomogeneity of the particle distribution produces deviations of the pdf $P\left(A/\langle A \rangle_{\bm{x}_i}\right)$ (the average being taken over all areas, containing each one particle) from the 
corresponding one computed for uniformly random distributed particles. 
As it can be seen in Fig.~\ref{fig:f8}, for $Ro=0$, $P\left(A/\langle A \rangle_{\bm{x}_i}\right)$ agrees 
with the probability distribution expected for uniformly spread particles in a 2D domain~\cite{FN2007},
$f_{2D}\left(A/\langle A \rangle_{\bm{x}_i}\right) = 343/15 \sqrt{7/(2\pi)} \left( A/\langle A \rangle_{\bm{x}_i} \right)^{5/2} \exp{ \left(-7/2 A/\langle A \rangle_{\bm{x}_i} \right)}$ 
(solid gray line in the figure). When the Rossby number increases, however, its left tail gets monotonically higher, indicating that the probability of finding particles at small distances, and hence to observe clustering, is larger. 
We can contrast the case of $Ro=0.075$ with one where we advect
particles by its geostrophic component only. As expected from particle transport in geostrophic turbulence~\cite{elmaidi93}, the pdf corresponding to uniformly distributed particles is recovered [case of $(Ro=0.075)_g$ in Fig.~\ref{fig:f8}], which further proves that this phenomenon is entirely due to the ageostrophic flow component.

Aiming to understand where particles accumulate, we first look at the fine-scale properties of clustering. 
The latter originate from the contraction of volumes in the phase space (here coinciding with the physical space) of the dissipative 
($\bm{\nabla} \cdot \bm{u}<0$) dynamical system of Eq.~(\ref{eq:motiontracers}). Consequently, after a transient, the Lagrangian dynamics take place on a fractal set.
A common quantitative indicator of clustering is the correlation dimension~\cite{GP1983}, $D_2$, 
of the dynamical attractor.
A decrease to values $D_2<d$, with $d$ the dimension of the physical space ($d=2$ in the present case), 
indicates an increased occurrence of small distances separating particle pairs.
This fractal dimension is defined as: 
\begin{equation}
 D_2 = \lim_{r_p\rightarrow 0} \frac{\log[C(r_p)]}{\log(r_p)},
\label{eq:D2}
\end{equation}
with the correlation sum 
$C(r_p)$ given by
\begin{equation}
C(r_p)=\lim_{N_p\to \infty }\frac{2}{N_p(N_p-1)}\sum_{i,j>i}^{N_p}\Theta (r_p-\left | {\bm{x}_i}-{\bm{x}_j} \right |) \nonumber,
\label{eq:corrsum}
\end{equation}
where $\Theta$ is the Heaviside step function, $\bm{x}_i$ and $\bm{x}_j$ are the positions of particles belonging to pair $(i,j)$, and the distance $\left | {\bm{x}_i}-{\bm{x}_j} \right |$ is the shortest one, after taking into account the $2\pi$-periodicity of the computational box.
Equation (\ref{eq:D2}) then means that, for small $r_p$, the probability to find particle pairs separated by a distance less than $r_p$ scales as $C(r_p) \sim r_p^{D_2}$.

\begin{figure}
    \includegraphics[scale=0.68]{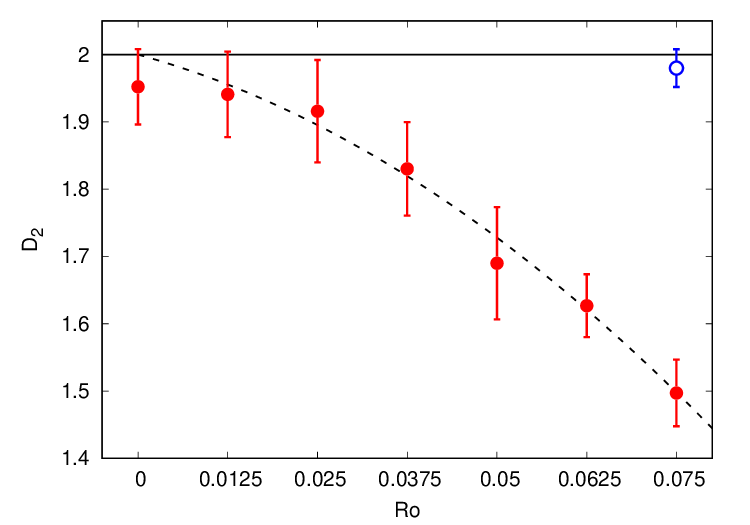}
    \caption{Correlation dimension $D_2$ as a function of $Ro$, 
    obtained from data in several statistically steady flow realizations. 
    Uncertainties are estimated from the standard deviations of best fits over the range 
    of small distances 
    $r_p$ where $C(r_p) \sim r_p^{D_2}$. 
    The empty blue point is for particles advected by the geostrophic flow component only at $Ro=0.075$. 
    The black dashed line corresponds to the second-order Taylor expansion $D_2 \simeq 2+aRo+bRo^2$, 
    with $a \simeq -2.9$ and $b \simeq -50.2$ from a fit.
    }
    \label{fig:f9}
\end{figure}

Figure~\ref{fig:f9} shows the measurement of the correlation dimension as a function of the Rossby number. 
For $Ro=0$, as expected, $D_2=2$ within statistical accuracy, 
which confirms the spatially homogeneous distribution of particles in the SQG system. 
Here, the small deviation from the theoretical value $2$ may be attributed to the finite number 
of particles.
At nonzero values of $Ro$, $D_2$ decreases monotonically, highlighting that clustering now takes place and that its intensity grows with the Rossby number.  
Again, this is a direct consequence of the transport of Lagrangian tracers by the ageostrophic flow. Indeed, when advection is realized by the geostrophic velocity only in the SQG$^{+1}$ model, the nonhomogeneity of the particle distribution disappears and $D_2 \simeq 2$, as shown by the blue empty point in the figure for the highest value of $Ro$ explored (but the same holds for all $Ro$). Overall, these results suggest that particles aggregate on flow structures with a dimensionality smaller than that of the physical space and progressively more unidimensional with increasing $Ro$.  

We now discuss in what regions of the flow particles tend to cluster. The question is of primary importance in oceanography, e.g. to identify areas of pollutant accumulation in surface flows, or locations of intense vertical velocities relevant for nutrient upwelling and plankton dynamics. 

While inspection of Fig.~\ref{fig:f3}d already suggests some tendency of particles to avoid negative-vorticity (anticyclonic) regions and to concentrate along filamentary structures, a more quantitative approach is needed. 
A classical tool to identify different (2D) flow regions, and to characterize their role 
in transport phenomena, is the Okubo-Weiss parameter~\cite{Okubo1970,Weiss1991},
\begin{equation}
Q = \sigma^2 - \zeta^2,
\label{eq:OW}
\end{equation}
where $\sigma=\sqrt{\sigma_n^2+\sigma_s^2}$ is the total strain ($\sigma_n = \partial_x u - \partial_y v$ 
and $\sigma_s = \partial_x v + \partial_y u$ being the normal and shear strain, respectively) and $\zeta$ is vorticity. 
The parameter $Q$ allows to discriminate between strain-dominated ($Q>0$, i.e. $\sigma>|\zeta|$) and rotation-dominated ($Q<0$, i.e. $\sigma<|\zeta|$) regions, and reveals useful, for instance, to explain the dynamics of tracer-field gradients~\cite{Lapeyre_etal00,klein2000}. 
Note that a more refined criterion was further obtained in incompressible flows to take into account the rotation of the strain eigenvectors that can affect the straining properties~\cite{Lapeyre_etal99}.
These strain and rotation-dominated regions can be related to dispersion properties through the linearization $d(\bm{x}_i-\bm{x}_j)/dt=\bm{u}_i-\bm{u}_j\simeq(\bm{\nabla} \bm{u})(\bm{x}_i-\bm{x}_j)$. It is then clear that velocity gradients will also determine the particle small-scale dispersion or aggregation properties. 

\begin{figure}
    \includegraphics[scale=0.54]{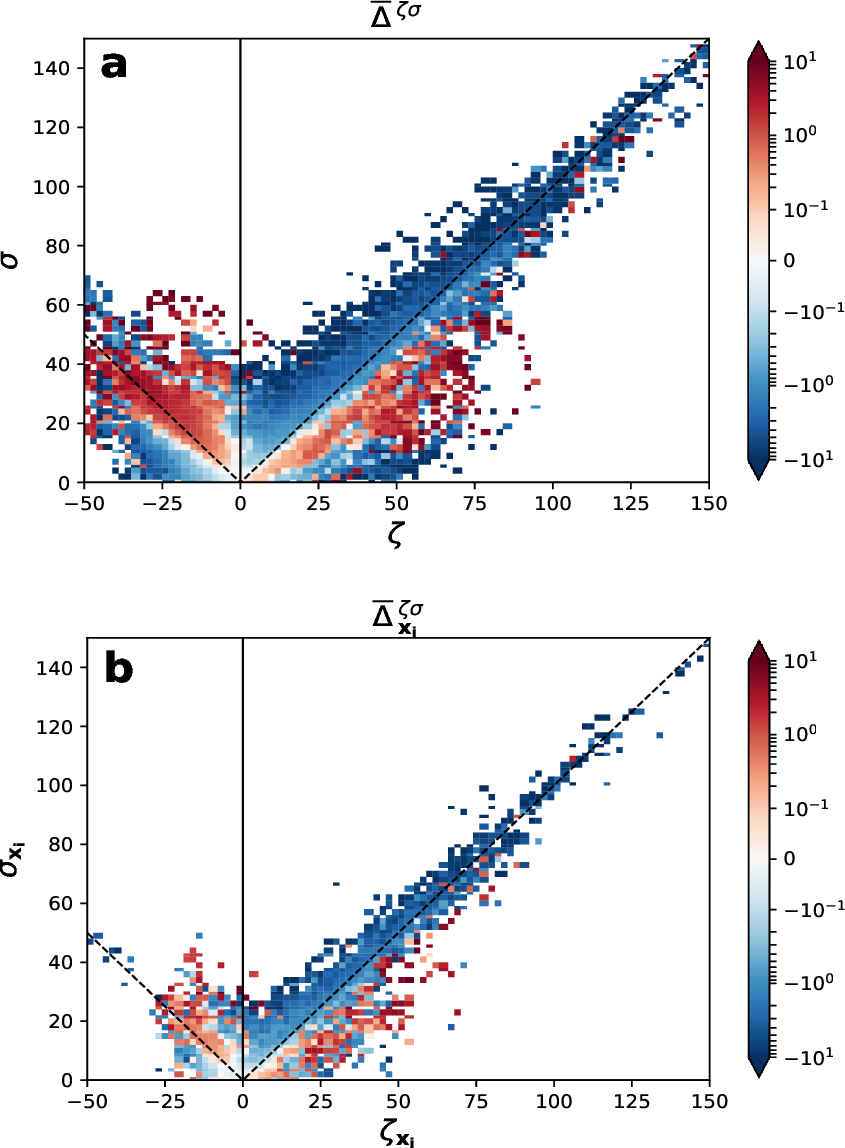}
    \caption{
    Mean divergence $\overline{\Delta}^{\zeta \sigma}$ 
    conditionally averaged over vorticity ($\zeta$) and strain ($\sigma$), from Eulerian (a) and Lagrangian (b) statistics, at a fixed instant of time in the statistically steady state of the flow, for $Ro=0.075$.
    For the Lagrangian estimate, the subscript $\bm{x}_i$ indicates that $\Delta$, $\zeta$ and $\sigma$ are computed at particle positions. In both (a) and (b) the dashed lines correspond to $\sigma=|\zeta|$.
    }
    \label{fig:f10}
\end{figure}

In order to determine the regions where particles preferentially cluster, we follow Ref.~\onlinecite{Balwada_etal_2021} and compute
the flow divergence conditionally averaged over all grid points of the domain with given values of vorticity and strain, noted $\overline{\Delta}^{\zeta \sigma}$.
This is a robust statistical tool originally introduced to investigate the vertical fluxes of a passive scalar field in submesoscale turbulence~\cite{Balwada_etal_2021}. Figure~\ref{fig:f10}a shows its measurement in our SQG$^{+1}$ simulations for $Ro=0.075$ at the same instant of time chosen for the visualization of Fig.~\ref{fig:f3}d (but it was verified that its features do not significantly change when a time-average is also taken).
It is here apparent that strong divergence ($\overline{\Delta}^{\zeta \sigma}>0$) and convergence ($\overline{\Delta}^{\zeta \sigma}<0$) predominantly occur in strain-dominated regions ($\sigma>|\zeta|$), extending along tails above the lines $\sigma=|\zeta|$. 
The asymmetric shape of the tails is a direct consequence of the dominance of cyclonic vorticity (see Fig.~\ref{fig:f2}), due to ageostrophic dynamics.  
Here, the association of convergence with $\zeta>0$ values  is arguably due to the same vortex-stretching effects that amplify cyclonic vorticity (Sec.~\ref{sec:vortstat}). 
Note, too, that in rotation-dominated regions ($|\zeta|>\sigma$), the divergence $\overline{\Delta}^{\zeta \sigma}$ is more likely to take both positive and negative values that tend to cancel out more. 
The above features are generic, and also appear at smaller values of $Ro$ (not shown), except that the tails associated with large positive and negative values of $\overline{\Delta}^{\zeta \sigma}$ become more symmetric, and divergence is smaller in absolute value,  when the Rossby number is decreased.  

To complete the picture, we also show in Fig.~\ref{fig:f10}b the divergence, in vorticity-strain space, computed at particle positions, $\overline{\Delta}^{\zeta \sigma}_{\bm{x}_i}$. The Rossby number and the instant of time are the same as in Fig.~\ref{fig:f10}a (and, again, we verified that averaging over time does not considerably modify the results).  
By comparing Fig.~\ref{fig:f10}a and Fig.~\ref{fig:f10}b, it is evident that the 
Lagrangian and Eulerian estimates of divergence, conditionally averaged over the values taken by vorticity and strain,
share the same general characteristics (similarly to what is found for vertical velocity in Ref.~\onlinecite{Wang_etal_2022}). 
The partial attenuation of extreme events when using Lagrangian statistics is likely due to the smaller sample. Apart from this, it can be noted that the patterns from the Lagrangian estimate are sharper and characterized by a reduced frequency of $\overline{\Delta}^{\zeta \sigma}>0$ events, in comparison with those from the Eulerian estimate.
This is due to the tendency of particles to aggregate in flow-convergence regions, and hence to predominantly sample negative values of divergence. 
Overall, Fig.~\ref{fig:f10}b confirms the preference of Lagrangian tracers to concentrate in regions of positive vorticity and large strain ($\sigma>|\zeta|$). This finding quite nicely matches the spatial organization of particles that 
is observed from a closeup view of a portion of the full domain at the same instant of time
(Fig.~{\protect\ref{fig:f3}}d).
Indeed, regions of negative vorticity ($\zeta<0$) tend to be relatively particle-free. 
On the contrary, particles are abundant in filamentary, positive vorticity regions (corresponding to $\zeta>0$ and $\sigma>\zeta$) while it is less the case inside cyclonic eddies (corresponding to $\zeta>0$ and $\sigma<\zeta$).

The previous analysis indicates that particle clustering takes place in cyclonic strain-dominated regions. 
These correspond mostly to filaments and fronts outside coherent eddies. 
Indeed, a straight front along the $y$ direction [with velocity $\bm{u}=\bm{u}(x)$ independent of $y$] is characterized by negative divergence 
($\bm{\nabla}\cdot\bm{u}=\partial_x u<0$) in its crosswise direction (which sustains the front) and by strain exceeding vorticity. 
The fact that $\sigma>|\zeta|$ follows from the relation $\sigma^2=(\bm{\nabla}\cdot\bm{u})^2+\zeta^2>\zeta^2$ holding for a velocity field that only depends on the cross-front coordinate $x$.

Our findings support those from a recent, more complex modeling study, which, taking an Eulerian point of view, reported on strong vertical velocities and flow convergence in cyclonic submesoscale fronts~\cite{Balwada_etal_2021}. 
Furthermore, they provide clear evidence of Lagrangian-tracer clustering in cyclonic regions, also observed from real surface-drifter data~\cite{Dasaro_etal_2018,vic2022}, and a possible explanation of the basic mechanisms controlling the phenomenon  in the framework of a minimal model accounting for ageostrophic dynamics.  

\section{Conclusions}
\label{sec:concl}

We studied Lagrangian particle dynamics in an idealized model of surface-ocean turbulence that includes ageostrophic motions by means of numerical simulations. We particularly focused on the effect of ageostrophy on the spreading 
process of tracer particles, by examining both relative dispersion and clustering properties. 

The turbulent dynamics were assumed to be described by the SQG$^{+1}$ system, which accounts for frontogenetic ageostrophic motions, and is obtained from a development of primitive equations to next order in $Ro$, with respect to standard QG models. 
This approach, originally introduced in an atmospheric context~\cite{HSM2002}, allowed us to reproduce the cyclone-anticyclone asymmetry, a phenomenon that is observed in both primitive-equation simulations~\cite{RK2010} and data from observations~\cite{Rudnick2001,Shcherbina_etal_2013} of ocean turbulence at sufficiently fine scales, but is missed by QG models. 
The turbulent flows from our simulations for different Rossby numbers are characterized by energetic small scales, particularly in the form of filamentary structures associated with intense gradients. 
Kinetic energy spectra are not far from the theoretical expectation in the SQG system (recovered by setting $Ro=0$ in the governing equations), although slightly steeper. 
Their scaling behavior is close to $E(k) \sim k^{-2}$,  as also found at submesoscales in more realistic simulations~\cite{Klein_etal_2008,Capet_etal_2008,BSA2018}. 
In the present case, the steepening of the spectrum is most likely due to the presence of large-scale coherent structures, a feature that was already observed in both the SQG~\cite{Lapeyre2017,CCMV2004} and the SQG$^{+1}$ systems~\cite{HSM2002}. 
 
To explore how ageostrophic fluid motions impact the particle separation process, we compared the measurements from different indicators of pair dispersion as a function of $Ro$. Given that the total kinetic energy increases when increasing $Ro$, 
we used mostly dimensionless diagnostics allowing
a fair comparison between the different simulations. We found that, irrespective of the Rossby number, dispersion is essentially nonlocal, except perhaps on a narrow range of separations, as highlighted by the extended region of scale independent FSLE and by the fast initial growth in time of the kurtosis of relative displacements. 
As the FSLE, where constant, was found to be close to the inverse large-eddy turnover time of the flow, we could show that this apparently surprising result is due to the presence of large persistent flow structures, which dominate the dispersion process.  
Overall, the general picture emerging from different metrics of relative dispersion is that, in the present simulations, dispersion only weakly depends on the intensity of the ageostrophic flow dynamics (i.e. $Ro$). 
Nevertheless, when increasing $Ro$, the latter manifest in a small, but measurable, increase of the separation rate at short times (and small distances), due to velocity gradients becoming stronger, and in a subsequent slowdown of relative dispersion at later times, possibly arising from the formation of temporary particle aggregations. 

The occurrence of clustering events was demonstrated by computing the averaged divergence experienced by particles (the dilation rate~\cite{HLJK2015}), and the pdf of cell areas from a Vorono\"i tessellation. The decrease of the dilation rate to more and more 
negative values, and the rise of the left tail of the Vorono\"i cell-area pdf, 
indicate that particles are progressively more likely to be at small distances one from the other, when $Ro$ is increased.  
While this phenomenon is a direct consequence of the compressibility of the ageostrophic flow component, it is not straightforward to relate Eulerian and Lagrangian 
measures of clustering, as already 
noted in previous studies of Lagrangian tracer dynamics in compressible turbulence~\cite{BDES2004,DMV2014}. 
Here, at a qualitative level, we argued that clustering arises from the increased probability of very large flow divergence values, at larger $Ro$, and hence the longer fraction of time spent by particles in 
negative-divergence regions.   

Determining where convergence, and thus particle clustering, takes place in surface-ocean flows is of paramount importance, both to predict the accumulation of biogeochemical substances or pollutants, and to identify locations of large vertical velocities. 
To address this question, we first computed the correlation dimension of the sets over which particles concentrate, which is directly related to the probability of finding a pair of them within a given distance. 
With increasing $Ro$, this was found to decrease from $D_2=2$ (corresponding to uniformly distributed particles) to smaller values, indicative of clustering and pointing to less than 2D aggregates (possibly quasi one-dimensional ones, 
for large enough Rossby numbers). To further understand in what flow regions clusters can be found, we examined the divergence conditionally averaged over vorticity and strain.
This quantity was recently introduced as a generalization of Okubo-Weiss parameter to divergent flows, in order to partition 2D flows into regions with different stirring properties~\cite{Balwada_etal_2021}. 
We found that divergence has an asymmetric distribution in vorticity-strain space that reflects the cyclone-anticyclone asymmetry. 
More interestingly, it is predominantly negative and large (in absolute value) where strain overcomes vorticity and the latter is positive, which indicates that clusters form in cyclonic frontal regions. 
Such a picture agrees with the results in more realistic simulations of submesoscale dynamics in the Antarctic Circumpolar Current, focused on the vertical fluxes of tracer fields~\cite{Balwada_etal_2021}. 
It may also be useful to better understand observations of surface-drifter clustering in cyclonic regions in the Gulf of Mexico~\cite{Dasaro_etal_2018}.   

To conclude, the SQG$^{+1}$ system revealed a useful minimal model to investigate some basic mechanisms, related to ageostrophy, 
controlling the separation and clustering of Lagrangian tracer particles at the ocean surface.
Ageostrophic effects only weakly affect the nonlocal relative dispersion while they are responsible of non-negligible clustering in filamentary cyclonic regions. 
This is remarkably similar to the observations from drifters in the Gulf of Mexico, which also indicated both nonlocal dispersion~\cite{Beron2016} and small-scale clustering~\cite{Dasaro_etal_2018}.
Note that, in addition to ageostrophy, in the real ocean, other processes play a role in the transport of particles in the surface layer, such as Ekman currents induced by the wind\cite{onink2019}, or Stokes drift due to ocean waves. The dispersion of floating material may also be affected by inertial effects\cite{beronvera_2019} or by the drag exerted by the wind (the so-called windage).
A natural perspective of this study is to extend the analysis to realistic simulations, in order to explore the effects of the ocean fast variability, which cannot be accounted for by the modeling framework considered here. 

Finally, the present results also appear to us interesting  in consideration of the satellite data at high spatial resolution 
acquired by the SWOT spatial mission~\cite{Morrow_etal_2019}. 
The weak dependence of pair-dispersion indicators on the Rossby number suggests that the geostrophically derived surface velocities may be essentially accurate for relative-dispersion applications. 
On the other hand, to access finer details of the particle dynamics, such as clustering phenomena, further information on the ageostrophic flow components would clearly be required. 

\begin{acknowledgments}
This work is a contribution to the joint CNES-NASA SWOT project DIEGO and is supported by the French CNES TOSCA program.
\end{acknowledgments}

\section*{Data availability}
The data that support the findings of this study are available from the corresponding author upon reasonable request.

\vspace{2mm}
\bibliography{references}

\end{document}